\documentclass[journal=jacsat,manuscript=article]{achemso}

\usepackage[version=3]{mhchem} 

\usepackage[version=3]{mhchem} 
\usepackage{physics}
\usepackage{amsmath, amssymb}
\usepackage{parskip} 
\usepackage[utf8]{inputenc}
\usepackage{subcaption}
\usepackage[toc,page]{appendix}
\usepackage{algorithm}
\usepackage{xcolor}
\usepackage{algpseudocode}
 \SectionNumbersOn

\author{Rishabh Gupta}
\affiliation[Purdue University]
{Department of Chemistry, Purdue University, West Lafayette, IN, USA}

\author{Manas Sajjan}
\affiliation[Purdue University]
{Department of Chemistry, Purdue University, West Lafayette, IN, USA}

\author{Raphael D. Levine}
\affiliation[Hebrew University]
{The Fritz Haber Center for Molecular Dynamics and Institute of Chemistry, The Hebrew University of Jerusalem, Jerusalem 91904, Israel
}
\alsoaffiliation[University of California]
{Department of Chemistry and Biochemistry and  Department of Molecular and
Medical Pharmacology, David Geffen School of Medicine, University of California, Los
Angeles, CA 90095, USA}

\author{Sabre Kais}
\email{kais@purdue.edu}
\affiliation[Purdue University]
{Department of Chemistry, Department of Physics and Astronomy, and Purdue Quantum Science and Engineering Institute, Purdue University, West Lafayette, IN, USA}

\title[An \textsf{achemso} demo]
{Variational Approach to Quantum State Tomography based on Maximal Entropy Formalism}

\setcitestyle{numbers}
\mciteErrorOnUnknownfalse

\begin{document}







\begin{abstract}
Quantum state tomography is an integral part of quantum computation and offers the starting point for the validation of various quantum devices. One of the central tasks in the field of state tomography is to reconstruct with high fidelity, the quantum states of a quantum system. From an experiment on a real quantum device, one can obtain the mean measurement values of different operators. With such a data as input, in this report we employ the maximal entropy formalism to construct the least biased mixed quantum state that is consistent with the given set of expectation values. Even though in principle, the reported formalism is quite general and should work for an arbitrary set of observables, in practice we shall demonstrate the efficacy of the algorithm on an informationally complete (IC) set of Hermitian operators. Such a set possesses the advantage of uniquely specifying a single quantum state from which the experimental measurements have been sampled and hence renders the rare opportunity to not only construct a least-biased quantum state but even replicate the exact state prepared experimentally within a preset tolerance. The primary workhorse of the algorithm is re-constructing an energy function which we designate as the effective Hamiltonian of the system, and parameterizing it with Lagrange multipliers, according to the formalism of maximal entropy. These parameters are thereafter optimized variationally so that the reconstructed quantum state of the system converges to the true quantum state within an error threshold. To this end, we employ a parameterized quantum circuit and a hybrid quantum-classical variational algorithm to obtain such a  target state making our recipe easily implementable on a near-term quantum device.

\end{abstract}

\section{Introduction}
The method of uniquely characterizing the quantum mechanical state of a quantum system based on a series of measurements of an informationally complete (IC) set of Hermitian operators is called quantum state tomography (QST) [\cite{nielson},\cite{kais},\cite{photonic},\cite{qubits},\cite{cramer}] and forms an important basis for testing and validating quantum devices. However, the traditional approaches to QST are being exhausted to its limits [\cite{10-qubit}] because of certain limitations that accompanies those approaches. Some of these limitations correspond to exponential scaling of the traditional QST techniques with system size, which in turn require exponential amounts of storage and processing power to carry out computations. Along with this, since we are in the era of noisy intermediate-scale quantum (NISQ) [\cite{preskill2018quantum}] devices, the fidelity of measurements is also a limiting factor for performing state tomography efficiently as noisy measurements can lead to low fidelity of the reconstructed quantum state [\cite{detector}]. Another challenging task within the domain of QST is to reconstruct high fidelity quantum states [\cite{aharonov2013guest}] that can be used as a starting point while addressing problems in the field of condensed-matter physics and also in the validation of quantum technologies [\cite{anshu2021sample}]. Several research approaches have already been proposed that attempts to address one or the other limitations and it has paved the way for further advancements in this field. Some of these tomographic techniques include maximum likelihood estimation (MLE) [\cite{hradil},\cite{baumgratz2013scalable}], Bayesian mean estimation (BME) [\cite{lukens2020practical},\cite{lukens2020bayesian}], quantum overlap tomography [\cite{overlap}], shadow tomography [\cite{aaronson},\cite{Huang2020}], neural network tomography [\cite{Torlai2018},\cite{Carrasquilla2019},\cite{Palmieri2020},\cite{xin2019local}], and others [\cite{wich},\cite{jaynes},\cite{katz1967principles},\cite{sajjan2021quantum}]. In our previous work we  proposed a method of QST based on the formalism of maximal entropy from an incomplete set of measurements  [\cite{rishabh},\cite{gupta2021convergence}]. With that motivation, this research is an attempt to address the challenge of quantum state preparation in the field of QST based on a variational approach that can be easily implemented on a near-term quantum device.     \newline
The maximal entropy formalism [\cite{jaynes},\cite{raphy},\cite{raphy2}] provides the most unbiased probability distribution, based on the maximization of the von Neumann entropy of the system, subject to the constraints of the problem  [\cite{wich},\cite{Buzek_1997},\cite{alhassid},\cite{dagan}]. As a natural consequence, when combined with the method of Lagrange multipliers it leads to an expression of density operator given by Eq. (\ref{rho2}), that can serve as an optimal candidate for variational Gibbs sampling [\cite{aharonov2013guest}]. Inspired by this, the current work focuses on reconstructing the quantum state of a system, represented by the quantum Gibbs state and based on the formalism of maximal entropy, from mean measurement values of IC set of Hermitian operators. Sampling from a probability distribution corresponding to quantum Gibbs state plays an important role in a variety of diverse fields within and not limited to many-body physics [\cite{lewin2021classical},\cite{anshu2021sample}], quantum simulations [\cite{childs2018toward}], quantum optimization [\cite{somma2008quantum}], and quantum machine learning [\cite{kieferova2017tomography},\cite{biamonte2017quantum}]. However, preparing Gibbs state of a given Hamiltonian at arbitrary low temperature is not an easy task [\cite{wang2021variational}] and various approaches have been proposed, both classical and quantum [\cite{poulin2009sampling},\cite{temme2011quantum},\cite{kastoryano2016quantum},\cite{brandao2019finite}], to prepare Gibbs state under certain specified conditions. Some of these techniques include algorithms based on quantum rejection sampling [\cite{yung2012quantum}], dynamics simulation [\cite{kaplan2017ground},\cite{riera2012thermalization}], dimension reduction [\cite{bilgin2010preparing}] but the overhead quantum resource cost of implementing these approaches is very high and not suitable for execution on near-term quantum devices. In order to find applications of quantum algorithms on NISQ devices the underlying quantum circuit should be shallow with low circuit depth and low number of qubits. Variational quantum algorithm (VQA) [\cite{mcclean2016theory}] is one such class of hybrid quantum-classical algorithm that follows a heuristic approach based on the variational principle and has been quite popular in the recent years  [\cite{bravo2019variational},\cite{huang2019near},\cite{larose2019variational},\cite{cerezo2020variational},\cite{peruzzo2014variational},\cite{chen2021variational}] owing to their implementation on NISQ devices with shallow quantum circuits. \newline
To prepare a quantum Gibbs state on a NISQ device using VQAs, several methods have been proposed [\cite{islam2015measuring},\cite{verdon2019quantum},\cite{wu2019variational},\cite{mcardle2019variational},\cite{yuan2019theory},\cite{chowdhury2020variational}]. In this work, we have employed the approach by Wang et al. [\cite{wang2021variational}] wherein the loss function for preparing the Gibbs state on the quantum circuit involves the truncation of the Taylor series for the entropy and is shown to prepare Gibbs state for a given Hamiltonian with fidelity of over 99$\%$. The physical Hamiltonian of the system is unknown and is in fact unnecessary in this protocol. One only has access to the expectation values of arbitrary set of Hermitian operators. In principle, using the formalism one can generate a least-biased quantum state consistent with such an arbitrary and even incomplete set of mean measurements, yet in this report we use an IC set for testing and validation with the hope of affording a near-exact re-construction of the unknown pure quantum state used for sampling. This is attained by constructing a Hermitian matrix \textbf{H}, parameterized by Lagrange multipliers. The latter serves as a proxy Hamiltonian for the construction of the Gibbs state that represents the tomographic reconstruction of the state of the quantum system. The methodology is elaborately discussed in Section \ref{max}. To validate the proposed approach, the formalism is implemented in IBM Qiskit [\cite{Qiskit}] and the results corresponding to the fidelity and trace distance between the reconstructed quantum state and the true state are shown in Section \ref{result}. 


\section{Methodology} \label{max}
\begin{figure}[ht!]
  \centering 
\includegraphics[width=6.5in]{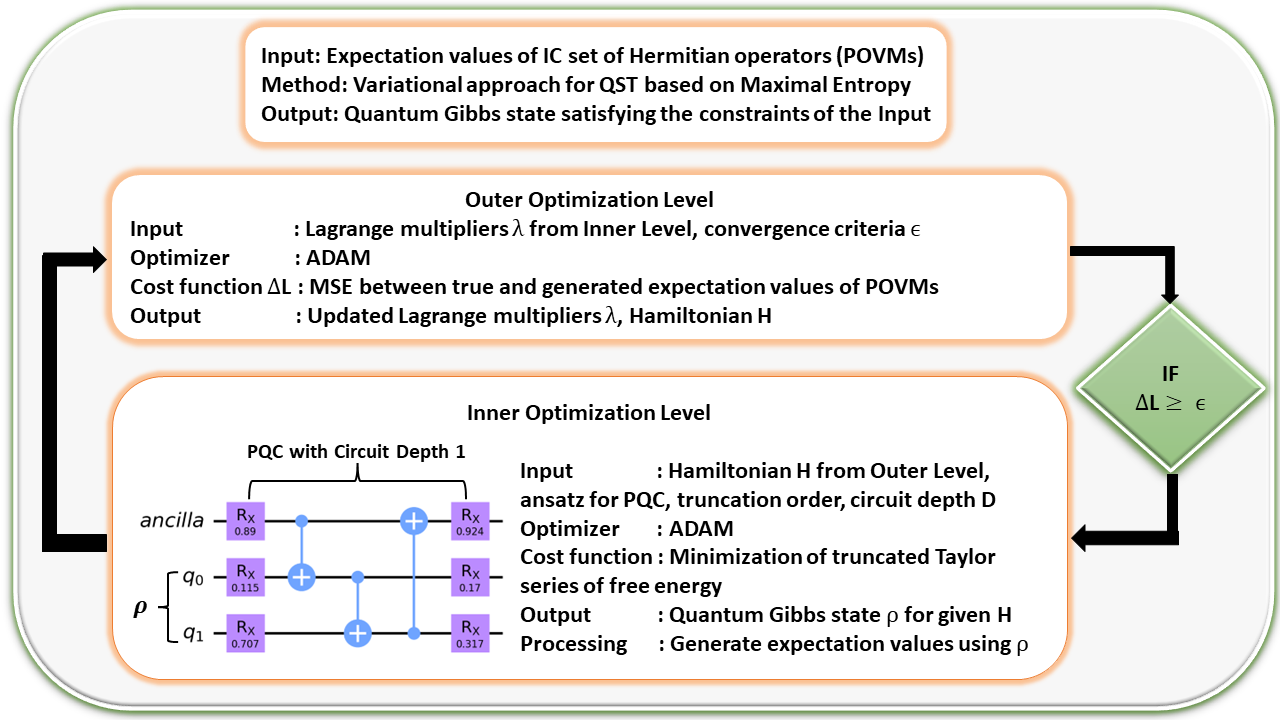}
\caption{The hybrid quantum-classical variational algorithm for quantum state tomography based on the formalism of maximal entropy.}
\label{fig_algo}
\end{figure}

The reconstruction of an unknown quantum state requires the information of a complete set of observables that are obtained through experimental measurements of Hermitian operators usually defined as positive-operator-valued measures (POVMs). The formalism of maximal entropy provides a unique characterization of the quantum state subject to the expectation values of a given set of operators that serve as the constraints of the problem. It also ensures that the Von Neumann entropy of the proposed distribution is maximum under the given constraints. The maximal entropy formalism when combined with the method of Lagrange multipliers $\lambda_k$ $\in$ $\mathbb{C}^2$ [\cite{jaynes},\cite{raphy2}], yields the following expression for the density operator of the unknown quantum state [\cite{wich},\cite{dagan}]:
\begin{eqnarray}
\hat{\rho} = \frac{1}{Z(\lambda_{1},\ldots,\lambda_{k})}\exp\{-\sum_{k}\lambda_{k}\hat{f}_{k}\} \label{rho2}
\end{eqnarray}
where $\hat{f}_k$ corresponds to the operators whose expectation values are known and $Z(\lambda_{1},\ldots,\lambda_{k})=Tr(\exp\{-\sum_{k}\lambda_{k}\hat{f}_{k}\})$ insures normalization as $Tr(\hat{\rho})=1$. The formalism of maximal entropy in general outputs a mixed state that is parameterized by the Lagrange multipliers as shown in Eq. (\ref{rho2}). In our approach since the target state is pure, these Lagrange multipliers are optimized such that the initial mixed state from the recipe gradually approaches idempotency during the training process and hence converges within an  $\epsilon$-neighborhood of the pure target state with $\epsilon$ being the error tolerance specified by the user. For example, consider the case a 2-qubit quantum system that can be uniquely described by the informationally complete (IC) set of Hermitian operators given by:
\begin{eqnarray}
    &\hspace*{0.001cm}&\{\ket{1}\bra{1},\ket{2}\bra{2},\ket{3}\bra{3},\ket{4}\bra{4},(\ket{1}\bra{2}\pm\ket{2}\bra{1}),(\ket{1}\bra{3}\pm\ket{3}\bra{1}),(\ket{1}\bra{4}\pm\ket{4}\bra{1}), \nonumber \\
    &\hspace*{0.001cm}&(\ket{2}\bra{3}\pm\ket{3}\bra{2}), (\ket{2}\bra{4}\pm\ket{4}\bra{2}), (\ket{3}\bra{4}\pm\ket{4}\bra{3}) \} \label{basis1}    
\end{eqnarray}
The expectation values of the initial four of these operators correspond to the probabilities and the rest are the coherences of the 2-qubit quantum system [\cite{alhassid}]. The IC set of these operators can be obtained using the linear combinations of Pauli string operators ($\sigma_x$, $\sigma_y$, $\sigma_z$, and $\sigma_i$) [\cite{kandala2017hardware},\cite{bian2019quantum},\cite{rishabh}]:
\begin{eqnarray}
    x_{11} = \langle\ket{1}\bra{1}\rangle &=& \frac{1}{4}(\sigma_z^2\sigma_z^1 + \sigma_z^2\sigma_i^1 + \sigma_i^2\sigma_z^1 + \sigma_i^2\sigma_i^1) \nonumber \\
    x_{12} = \langle\ket{1}\bra{2}+\ket{2}\bra{1}\rangle &=& \frac{1}{2}(\sigma_z^2\sigma_x^1 +  \sigma_i^2\sigma_x^1) \nonumber \\    
    x_{22} = \langle\ket{2}\bra{2}\rangle &=& \frac{1}{4}(\sigma_z^2\sigma_z^1 - \sigma_z^2\sigma_i^1 + \sigma_i^2\sigma_z^1 - \sigma_i^2\sigma_i^1) \nonumber    
\end{eqnarray}
and so on. \newline
Analogous to the maximal entropy formalism, parameterized by a single parameter $\beta$ = 1/$k_BT$ where $k_B$ is the Boltzmann's constant and \textit{T} is the temperature, the quantum Gibbs state for a given Hamiltonian H is defined as:
\begin{eqnarray}
    \hat{\rho} = \frac{\exp{-\beta H}}{tr(\exp{-\beta H})} \label{gibbs}
\end{eqnarray}
Constructing the Gibbs state of a given Hamiltonian on a parameterized quantum circuit requires minimization of the Helmholtz free energy described by the function:
\begin{eqnarray}
    \emph{F}(\rho) = tr(\rho H) - \beta^{-1}S(\rho)
\end{eqnarray}
where \textit{S($\rho)$} = -tr($\rho$ln$\rho$) corresponds to the von Neumann entropy of $\rho$. However, the most challenging part of constructing the loss function that minimizes the free energy of the Hamiltonian is estimating the entropy of the parameterized quantum state [\cite{gheorghiu2020estimating}]. In this work, to address the problem we adopted the method proposed by Wang \textit{et al} [\cite{wang2021variational}] wherein they used the Taylor series of entropy and truncate it at order \textit{K} and therefore, the truncated free energy is set as the loss function of the variational quantum algorithm. This method is practical in its implementation on a near-term quantum device as essentially the loss function involves estimating higher-order state overlaps, tr($\rho^k$), that corresponds to the truncated von Neumann entropy and can be carried out on a quantum device using swap tests [\cite{buhrman2001quantum},\cite{gottesman2001quantum},\cite{islam2015measuring},\cite{patel2016quantum},\cite{linke2018measuring}]. \newline
The core idea of the current research work stems from the combination of Eq. (\ref{rho2}) and (\ref{gibbs}) as in Eq. (\ref{rho2}) we are interested in maximizing the entropy by optimizing the unknown Lagrange multipliers $\lambda_k$ according to the constraints of the expectation values of the set of Hermitian operators and in Eq. (\ref{gibbs}) we want to optimize the parameters of the quantum circuit in order to minimize the free energy that yields the quantum Gibbs state for a particular Hamiltonian. In our methodology, the exponent term in Eq. (\ref{rho2}) is a Hermitian matrix \textbf{H} that constitutes the Hamiltonian for which the Gibbs state given by Eq. (\ref{gibbs}) is constructed on a quantum circuit using a variational algorithm. Thus, the hybrid variational quantum algorithm that is employed basically involves two levels of optimization and is termed as inner and outer optimization levels as shown in Figure \ref{fig_algo}. For a fixed set of Lagrange multipliers $\lambda_k$, the Hamiltonian \textbf{H} is constructed and passed onto the inner optimization level where the circuit parameters are variationally optimized using the truncated free energy as the loss function to yield a quantum Gibbs state corresponding to \textbf{H}. The constructed Gibbs state is then sent to the outer optimization level where the expectation values of the set of Hermitian operators are computed using the generated Gibbs state and then the Lagrange multipliers $\lambda_k$ are updated so as to minimize the mean square error between the generated and true expectation values of the POVMs. The updated Lagrange multipliers yield the new proxy Hamiltonian \textbf{H} that is again sent back to the inner optimization level and this process continues until convergence of the generated expectation values to the true values. This variational approach to QST based on maximal entropy is theoretically generalizable to any number of qubits. \newline
The variational quantum circuit comprises of \textit{n}-qubits that corresponds to the size of the quantum system and also an additional ancilla qubit. The circuit incorporates a series of parameterized single qubit rotational gates on every qubit and each qubit is entangled to the next qubit using the controlled-NOT (CNOT) gates. This sequence of rotation and CNOT gates is repeated depending on the expressivity that is required to obtain high fidelity of the prepared quantum states. The scaling of the algorithm in terms of quantum resource allocation is strictly polynomial as to re-construct a generic \textit{n}-qubit pure quantum state, we require \textit{n}+1 qubits and $\mathcal{O}$(D\textit{n}) quantum gates wherein $D$ is the depth (number of repeating layers) of the circuit ansatz used. \newline
Using the aforesaid procedure, in this work, we were able to obtain a high fidelity of 0.99 by setting $D=2$ for two qubits system and $D=6$ for the six qubits quantum system. The choice of the single qubit rotational gates depends on whether the reconstructed state needs to be real or complex. In case of real quantum states, parameterized R$_y$ gates can be used else for generalized complex states one can choose R$_x$ gates in the sequence. 

\section{Results and discussion} \label{result}
\begin{figure}[ht!]
  \centering 
\includegraphics[width=3.2in]{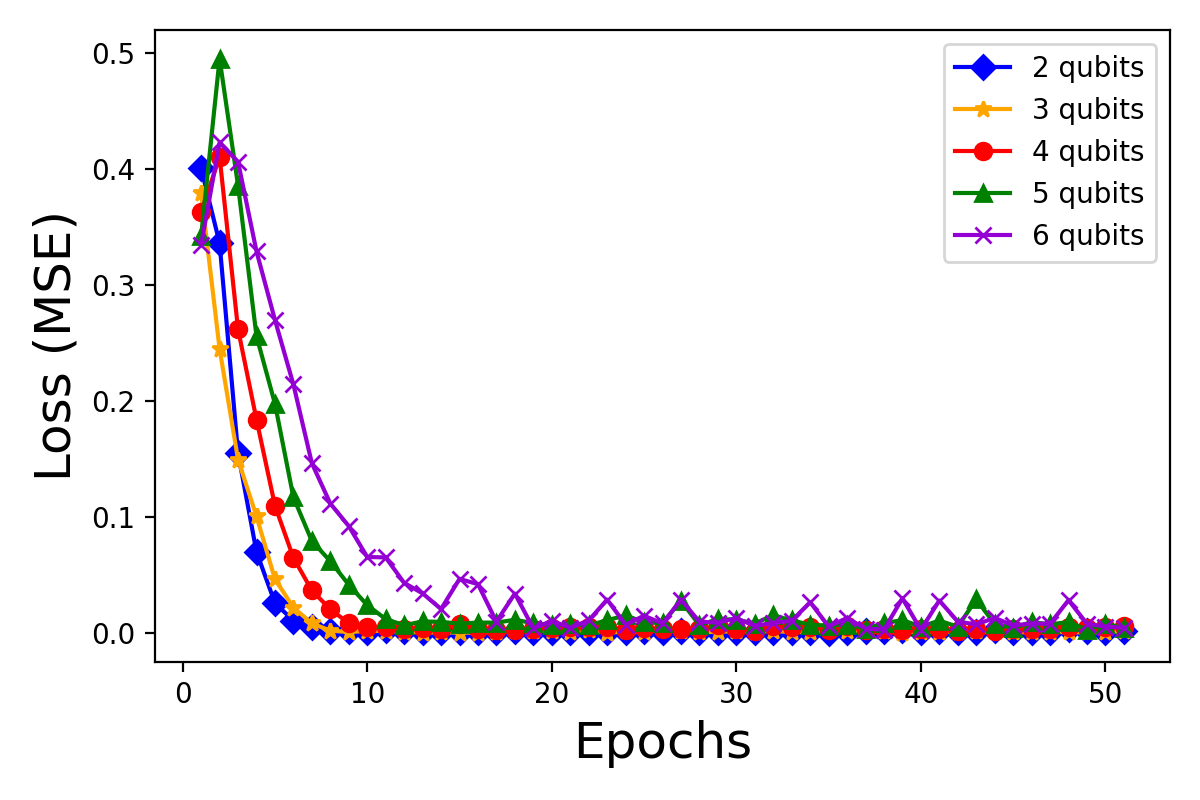} 
\caption{For the considered quantum systems with qubits ranging from 2-6, this plot shows the mean square error (MSE) loss as a function of the number of epochs between the true expectation values and the generated expectation values, of the IC set of Hermitian operators, obtained using the reconstructed quantum state upon each step of optimization. }
\label{fig_mse}
\end{figure}

\begin{figure}[ht!]
  \centering 
\includegraphics[width=3.2in]{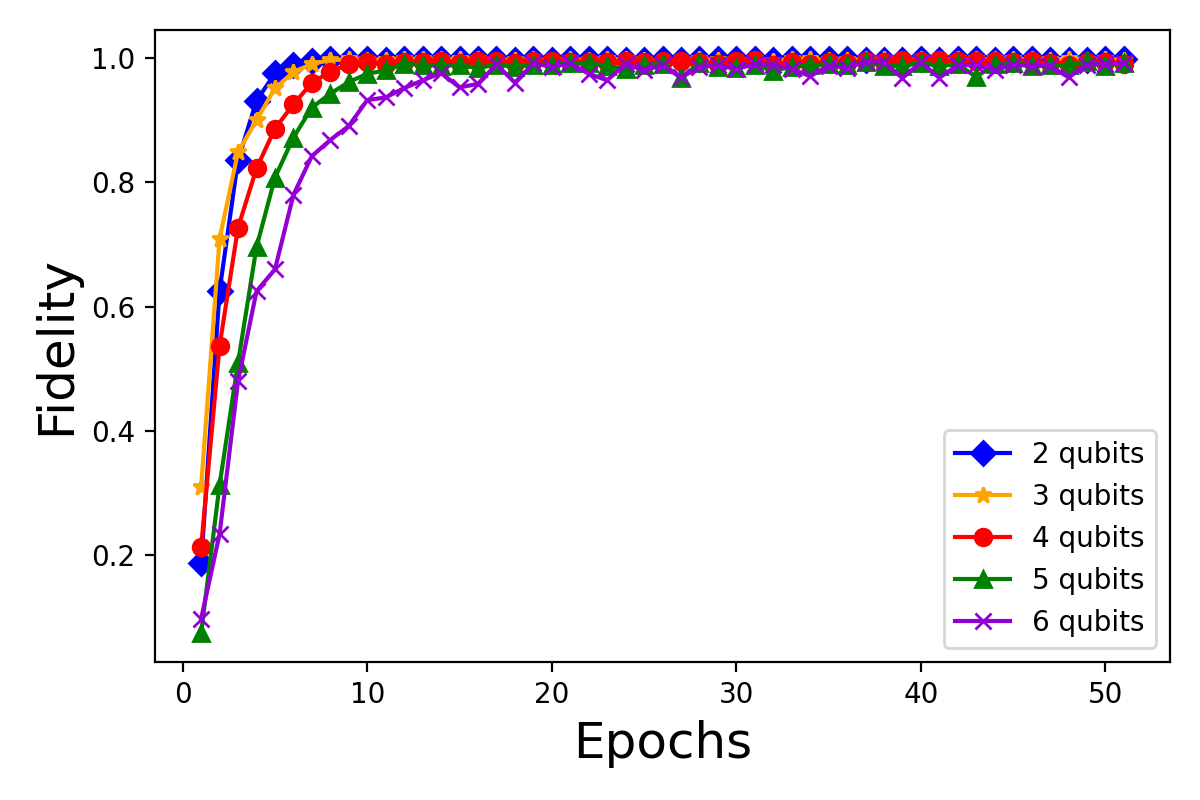} 
\caption{As a measure of convergence between the true and the reconstructed state, the fidelity between the two states is plotted as a function of the number of epochs. The fidelity close to 1 at the end of the optimization cycle shows that the proposed method based on maximal entropy formalism is able to successfully reconstruct the quantum state using the experimental mean measurement values of IC set of operators.}
\label{fig_fid}
\end{figure}

\begin{figure}[ht!]
  \centering 
\includegraphics[width=3.2in]{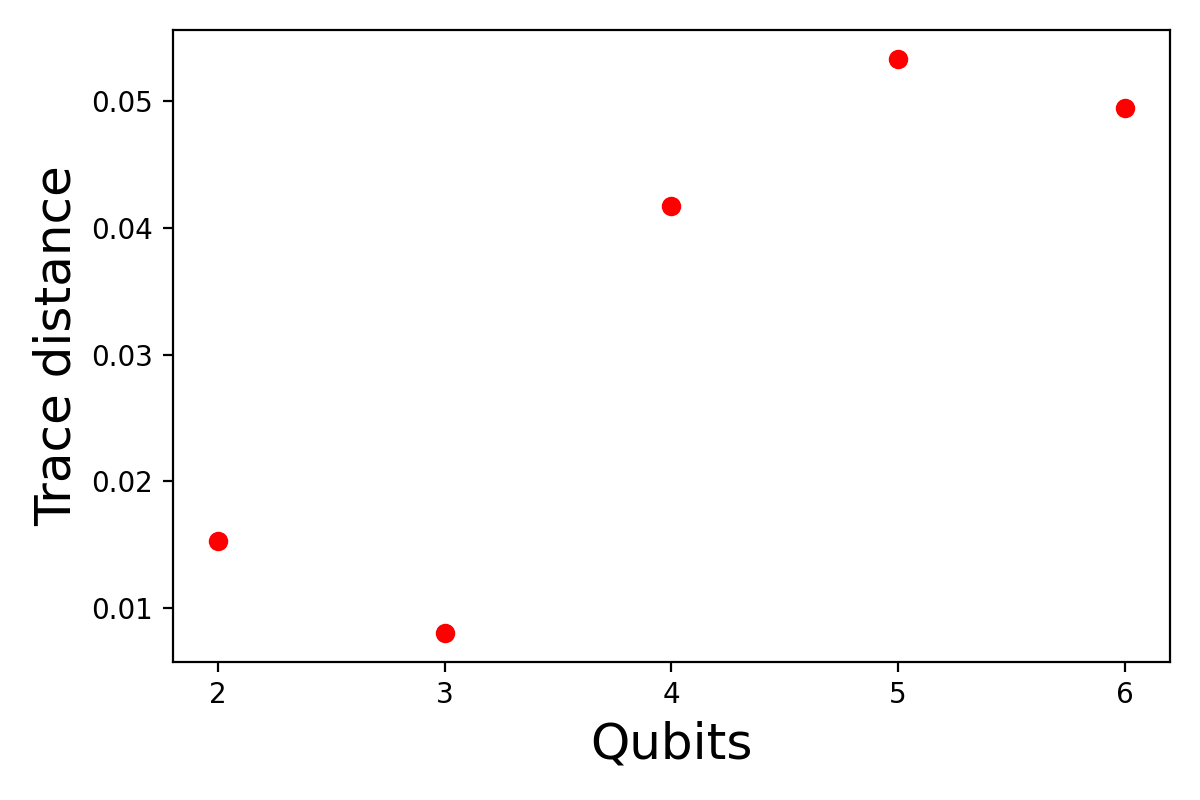} 
\caption{Trace distance between the true state and the final reconstructed quantum state obtained using the converged Lagrange multipliers at the end of the optimization process.}
\label{fig_trace}
\end{figure}

In this current work we propose a variational approach based on maximal entropy formalism to perform quantum state tomography on a near-term quantum device. This procedure outcomes a reconstructed quantum state that is prepared on a parameterized quantum circuit by using the expectation values of an IC set of Hermitian operators as input. The approach is tested and validated through numerical simulations conducted on IBM's Qiskit [\cite{Qiskit}] using its prototype quantum simulators for quantum systems consisting of up to 6 qubits. There are various backends available in Qiskit and we used the noise-free \textit{statevector$\_$simulator} backend to corroborate the theory. \newline
\begin{figure}[ht!]
  \centering 
\includegraphics[width=3.2in]{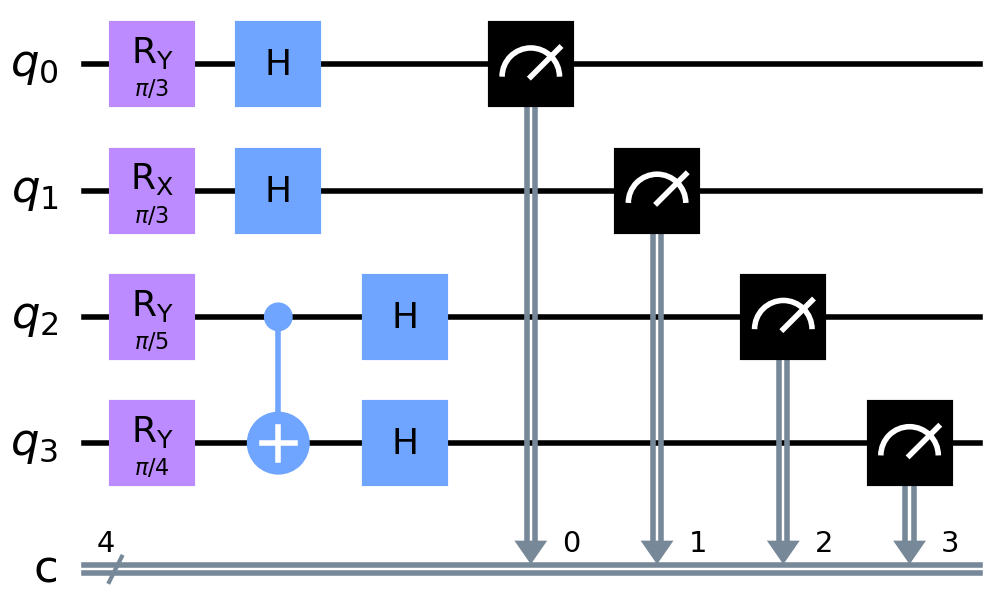} 
\includegraphics[width=3.2in]{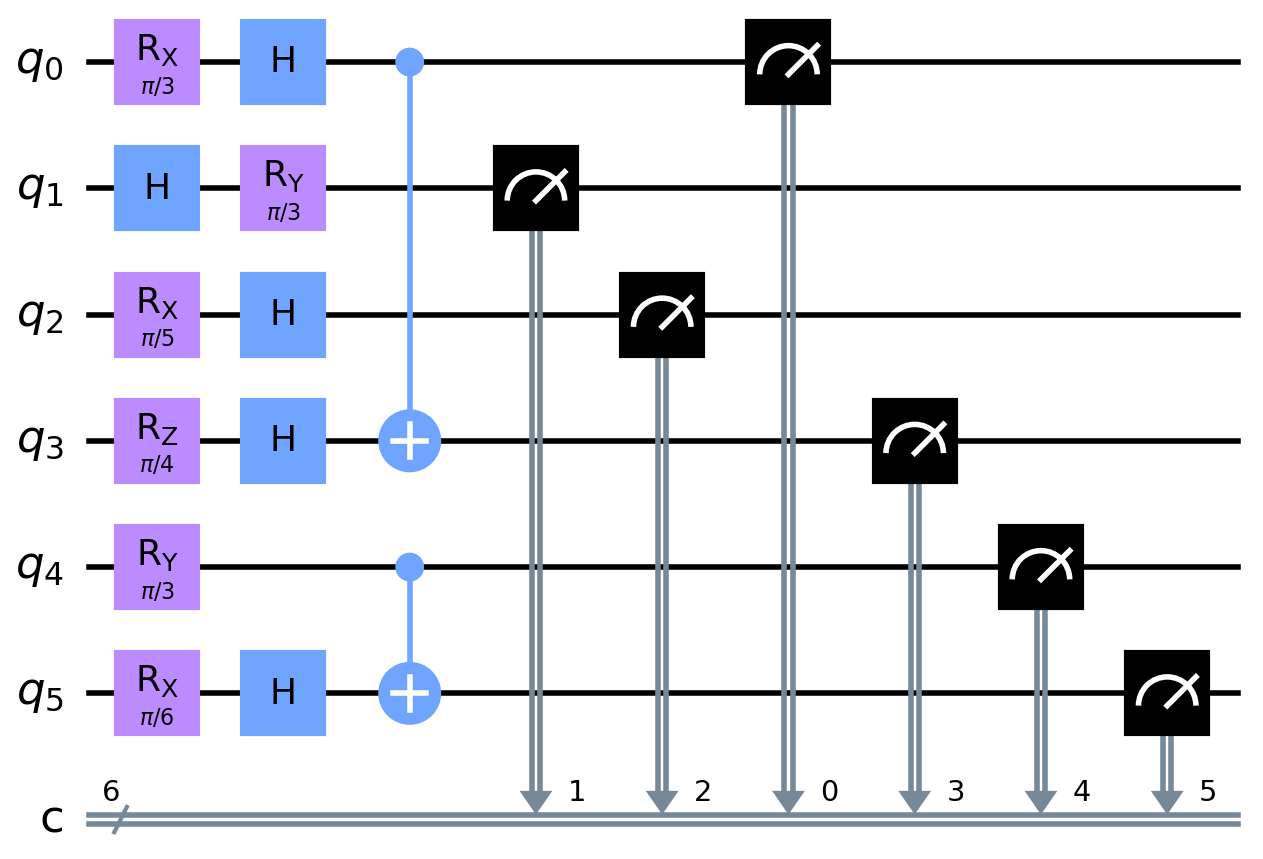} 
\caption{Sample 4-qubit and 6-qubit quantum circuits for which the quantum states are reconstructed using the proposed approach based on maximal entropy formalism.}
\label{fig_cir}
\end{figure} 
Different quantum circuits ranging from 2-6 qubits and consisting of one and two qubit quantum gates such as rotational, Hadamard, CNOT gates, etc. are used to prepare the sample states whose measurement statistics are reproduced using the reconstructed quantum state from the proposed maximal entropy based variational approach. Sample 4-qubit and 6-qubit quantum circuits are shown in Figure \ref{fig_cir}. As discussed in Section \ref{max}, at the end of full execution of inner optimization level a quantum Gibbs state is generated that is used to calculate the expectation values of the considered Hermitian operators. The mean square error (MSE) between the generated and the true expectation values serve as the loss function for updating the Lagrange multipliers in the outer optimization level. The MSE loss is plotted as a function of the epochs in Figure \ref{fig_mse}. As can be seen in Figure \ref{fig_mse}, the MSE loss converges to zero faster for smaller systems and the convergence becomes more erratic as the size of the quantum system increases. \newline
To perform quantum state tomography it is imperative that the reconstructed quantum state should be in close agreement with the true state. To demonstrate this a plot of fidelity between the reconstructed quantum state and the true state as a function of the number of epochs is shown in Figure \ref{fig_fid} for the different quantum systems with varying number of qubits. The figure shows the convergence of the reconstructed state to the true state as the Lagrange multipliers are updated with each step and the fidelity approaches 1 near the end of the optimization cycle. Another measure to test the performance of our approach is the trace distance between the true state ($\sigma$) and the reconstructed quantum state ($\rho$). Trace distance given by T($\rho,\sigma$) = $\frac{1}{2}||\rho-\sigma||_1$ is a measure of the closeness between two states. Figure \ref{fig_trace} shows the trace distance between the reconstructed quantum state, obtained from the converged set of $\lambda$ parameters, and the true state and as can be seen the trace distance for all the quantum systems considered is below 0.05 that also validates the successful reconstruction of the quantum state.

\section{Concluding remarks}
In this research we have shown the successful reconstruction of the quantum state based on a variational approach to QST by utilizing the formalism of maximal entropy and can be easily implemented on a near-term quantum device. The reconstructed state can reproduce the measurement statistics of the IC set of Hermitian operators that are considered while formulating the cost function for updating the variational parameters of the Hamiltonian arising from maximal entropy. The fidelity between the converged state and the true state is over 0.99 for all the considered quantum systems and the trace distance is below 0.05 that depicts the validation of the proposed approach for QST. This proposed variational approach can be applicable to a variety of different directions where QST is essential. For example, we intend to further study this approach for analyzing, characterizing, and mitigating single and double qubit quantum gates' errors while running computations on prototype quantum devices. This method can also be used as an efficient approach for sampling Gibbs state and therefore, can serve as a starting point for preparing targeted quantum state for quantum simulations. 

\begin{acknowledgement}
We acknowledge the financial support from the National Science Foundation under Award No. 1955907. We also like to acknowledge the partial financial support by the U.S. Department of Energy (Office of Basic Energy Sciences) under Award No. DE-SC0019215. We also acknowledge use of the IBM Q for this work. The views expressed here are those of the authors and do not reflect the official policy or position of IBM or the IBM Q team.

\end{acknowledgement}




\bibliography{achemso-demo}

\providecommand{\latin}[1]{#1}
\makeatletter
\providecommand{\doi}
  {\begingroup\let\do\@makeother\dospecials
  \catcode`\{=1 \catcode`\}=2 \doi@aux}
\providecommand{\doi@aux}[1]{\endgroup\texttt{#1}}
\makeatother
\providecommand*\mcitethebibliography{\thebibliography}
\csname @ifundefined\endcsname{endmcitethebibliography}
  {\let\endmcitethebibliography\endthebibliography}{}
\begin{mcitethebibliography}{68}
\providecommand*\natexlab[1]{#1}
\providecommand*\mciteSetBstSublistMode[1]{}
\providecommand*\mciteSetBstMaxWidthForm[2]{}
\providecommand*\mciteBstWouldAddEndPuncttrue
  {\def\EndOfBibitem{\unskip.}}
\providecommand*\mciteBstWouldAddEndPunctfalse
  {\let\EndOfBibitem\relax}
\providecommand*\mciteSetBstMidEndSepPunct[3]{}
\providecommand*\mciteSetBstSublistLabelBeginEnd[3]{}
\providecommand*\EndOfBibitem{}
\mciteSetBstSublistMode{f}
\mciteSetBstMaxWidthForm{subitem}{(\alph{mcitesubitemcount})}
\mciteSetBstSublistLabelBeginEnd
  {\mcitemaxwidthsubitemform\space}
  {\relax}
  {\relax}

\bibitem[Nielsen and Chuang(2011)Nielsen, and Chuang]{nielson}
Nielsen,~M.~A.; Chuang,~I.~L. \emph{Quantum Computation and Quantum
  Information: 10th Anniversary Edition}, 10th ed.; Cambridge University Press:
  USA, 2011\relax
\mciteBstWouldAddEndPuncttrue
\mciteSetBstMidEndSepPunct{\mcitedefaultmidpunct}
{\mcitedefaultendpunct}{\mcitedefaultseppunct}\relax
\EndOfBibitem
\bibitem[Kais(2014)]{kais}
Kais,~S. \emph{Quantum Information and Computation for Chemistry}; Wiley and
  Sons: Hoboken: NJ, 2014; Vol. 154\relax
\mciteBstWouldAddEndPuncttrue
\mciteSetBstMidEndSepPunct{\mcitedefaultmidpunct}
{\mcitedefaultendpunct}{\mcitedefaultseppunct}\relax
\EndOfBibitem
\bibitem[Altepeter \latin{et~al.}(2005)Altepeter, Jeffrey, and Kwiat]{photonic}
Altepeter,~J.; Jeffrey,~E.; Kwiat,~P. In \emph{Photonic State Tomography};
  Berman,~P., Lin,~C., Eds.; Advances In Atomic, Molecular, and Optical
  Physics; Academic Press, 2005; Vol.~52; pp 105 -- 159\relax
\mciteBstWouldAddEndPuncttrue
\mciteSetBstMidEndSepPunct{\mcitedefaultmidpunct}
{\mcitedefaultendpunct}{\mcitedefaultseppunct}\relax
\EndOfBibitem
\bibitem[James \latin{et~al.}(2001)James, Kwiat, Munro, and White]{qubits}
James,~D. F.~V.; Kwiat,~P.~G.; Munro,~W.~J.; White,~A.~G. Measurement of
  qubits. \emph{Phys. Rev. A} \textbf{2001}, \emph{64}, 052312\relax
\mciteBstWouldAddEndPuncttrue
\mciteSetBstMidEndSepPunct{\mcitedefaultmidpunct}
{\mcitedefaultendpunct}{\mcitedefaultseppunct}\relax
\EndOfBibitem
\bibitem[Banaszek \latin{et~al.}(2013)Banaszek, Cramer, and Gross]{cramer}
Banaszek,~K.; Cramer,~M.; Gross,~D. Focus on quantum tomography. \emph{New J.
  Phys.} \textbf{2013}, \emph{15}, 125020\relax
\mciteBstWouldAddEndPuncttrue
\mciteSetBstMidEndSepPunct{\mcitedefaultmidpunct}
{\mcitedefaultendpunct}{\mcitedefaultseppunct}\relax
\EndOfBibitem
\bibitem[Song \latin{et~al.}(2017)Song, Xu, Liu, Yang, Zheng, Deng, Xie, Huang,
  Guo, Zhang, Zhang, Xu, Zheng, Zhu, Wang, Chen, Lu, Han, and Pan]{10-qubit}
Song,~C. \latin{et~al.}  10-Qubit Entanglement and Parallel Logic Operations
  with a Superconducting Circuit. \emph{Phys. Rev. Lett.} \textbf{2017},
  \emph{119}, 180511\relax
\mciteBstWouldAddEndPuncttrue
\mciteSetBstMidEndSepPunct{\mcitedefaultmidpunct}
{\mcitedefaultendpunct}{\mcitedefaultseppunct}\relax
\EndOfBibitem
\bibitem[Preskill(2018)]{preskill2018quantum}
Preskill,~J. Quantum Computing in the NISQ era and beyond. \emph{Quantum}
  \textbf{2018}, \emph{2}, 79\relax
\mciteBstWouldAddEndPuncttrue
\mciteSetBstMidEndSepPunct{\mcitedefaultmidpunct}
{\mcitedefaultendpunct}{\mcitedefaultseppunct}\relax
\EndOfBibitem
\bibitem[Chen \latin{et~al.}(2019)Chen, Farahzad, Yoo, and Wei]{detector}
Chen,~Y.; Farahzad,~M.; Yoo,~S.; Wei,~T.-C. Detector tomography on IBM quantum
  computers and mitigation of an imperfect measurement. \emph{Phys. Rev. A}
  \textbf{2019}, \emph{100}, 052315\relax
\mciteBstWouldAddEndPuncttrue
\mciteSetBstMidEndSepPunct{\mcitedefaultmidpunct}
{\mcitedefaultendpunct}{\mcitedefaultseppunct}\relax
\EndOfBibitem
\bibitem[Aharonov \latin{et~al.}(2013)Aharonov, Arad, and
  Vidick]{aharonov2013guest}
Aharonov,~D.; Arad,~I.; Vidick,~T. Guest column: the quantum PCP conjecture.
  \emph{Acm sigact news} \textbf{2013}, \emph{44}, 47--79\relax
\mciteBstWouldAddEndPuncttrue
\mciteSetBstMidEndSepPunct{\mcitedefaultmidpunct}
{\mcitedefaultendpunct}{\mcitedefaultseppunct}\relax
\EndOfBibitem
\bibitem[Anshu \latin{et~al.}(2021)Anshu, Arunachalam, Kuwahara, and
  Soleimanifar]{anshu2021sample}
Anshu,~A.; Arunachalam,~S.; Kuwahara,~T.; Soleimanifar,~M. Sample-efficient
  learning of interacting quantum systems. \emph{Nature Physics} \textbf{2021},
  \emph{17}, 931--935\relax
\mciteBstWouldAddEndPuncttrue
\mciteSetBstMidEndSepPunct{\mcitedefaultmidpunct}
{\mcitedefaultendpunct}{\mcitedefaultseppunct}\relax
\EndOfBibitem
\bibitem[Hradil(1997)]{hradil}
Hradil,~Z. Quantum-state estimation. \emph{Phys. Rev. A} \textbf{1997},
  \emph{55}, R1561--R1564\relax
\mciteBstWouldAddEndPuncttrue
\mciteSetBstMidEndSepPunct{\mcitedefaultmidpunct}
{\mcitedefaultendpunct}{\mcitedefaultseppunct}\relax
\EndOfBibitem
\bibitem[Baumgratz \latin{et~al.}(2013)Baumgratz, N{\"u}{\ss}eler, Cramer, and
  Plenio]{baumgratz2013scalable}
Baumgratz,~T.; N{\"u}{\ss}eler,~A.; Cramer,~M.; Plenio,~M.~B. A scalable
  maximum likelihood method for quantum state tomography. \emph{New Journal of
  Physics} \textbf{2013}, \emph{15}, 125004\relax
\mciteBstWouldAddEndPuncttrue
\mciteSetBstMidEndSepPunct{\mcitedefaultmidpunct}
{\mcitedefaultendpunct}{\mcitedefaultseppunct}\relax
\EndOfBibitem
\bibitem[Lukens \latin{et~al.}(2020)Lukens, Law, Jasra, and
  Lougovski]{lukens2020practical}
Lukens,~J.~M.; Law,~K.~J.; Jasra,~A.; Lougovski,~P. A practical and efficient
  approach for Bayesian quantum state estimation. \emph{New Journal of Physics}
  \textbf{2020}, \emph{22}, 063038\relax
\mciteBstWouldAddEndPuncttrue
\mciteSetBstMidEndSepPunct{\mcitedefaultmidpunct}
{\mcitedefaultendpunct}{\mcitedefaultseppunct}\relax
\EndOfBibitem
\bibitem[Lukens \latin{et~al.}(2020)Lukens, Law, and
  Bennink]{lukens2020bayesian}
Lukens,~J.~M.; Law,~K.~J.; Bennink,~R.~S. A Bayesian analysis of classical
  shadows. \emph{arXiv preprint arXiv:2012.08997} \textbf{2020}, \relax
\mciteBstWouldAddEndPunctfalse
\mciteSetBstMidEndSepPunct{\mcitedefaultmidpunct}
{}{\mcitedefaultseppunct}\relax
\EndOfBibitem
\bibitem[Cotler and Wilczek(2020)Cotler, and Wilczek]{overlap}
Cotler,~J.; Wilczek,~F. Quantum Overlapping Tomography. \emph{Phys. Rev. Lett.}
  \textbf{2020}, \emph{124}, 100401\relax
\mciteBstWouldAddEndPuncttrue
\mciteSetBstMidEndSepPunct{\mcitedefaultmidpunct}
{\mcitedefaultendpunct}{\mcitedefaultseppunct}\relax
\EndOfBibitem
\bibitem[Aaronson(2018)]{aaronson}
Aaronson,~S. Shadow Tomography of Quantum States. Proceedings of the 50th
  Annual ACM SIGACT Symposium on Theory of Computing. New York, NY, USA, 2018;
  p 325–338\relax
\mciteBstWouldAddEndPuncttrue
\mciteSetBstMidEndSepPunct{\mcitedefaultmidpunct}
{\mcitedefaultendpunct}{\mcitedefaultseppunct}\relax
\EndOfBibitem
\bibitem[Huang \latin{et~al.}(2020)Huang, Kueng, and Preskill]{Huang2020}
Huang,~H.-Y.; Kueng,~R.; Preskill,~J. Predicting many properties of a quantum
  system from very few measurements. \emph{Nature Physics} \textbf{2020},
  \emph{16}, 1050--1057\relax
\mciteBstWouldAddEndPuncttrue
\mciteSetBstMidEndSepPunct{\mcitedefaultmidpunct}
{\mcitedefaultendpunct}{\mcitedefaultseppunct}\relax
\EndOfBibitem
\bibitem[Torlai \latin{et~al.}(2018)Torlai, Mazzola, Carrasquilla, Troyer,
  Melko, and Carleo]{Torlai2018}
Torlai,~G.; Mazzola,~G.; Carrasquilla,~J.; Troyer,~M.; Melko,~R.; Carleo,~G.
  Neural-network quantum state tomography. \emph{Nature Physics} \textbf{2018},
  \emph{14}, 447--450\relax
\mciteBstWouldAddEndPuncttrue
\mciteSetBstMidEndSepPunct{\mcitedefaultmidpunct}
{\mcitedefaultendpunct}{\mcitedefaultseppunct}\relax
\EndOfBibitem
\bibitem[Carrasquilla \latin{et~al.}(2019)Carrasquilla, Torlai, Melko, and
  Aolita]{Carrasquilla2019}
Carrasquilla,~J.; Torlai,~G.; Melko,~R.~G.; Aolita,~L. Reconstructing quantum
  states with generative models. \emph{Nature Machine Intelligence}
  \textbf{2019}, \emph{1}, 155--161\relax
\mciteBstWouldAddEndPuncttrue
\mciteSetBstMidEndSepPunct{\mcitedefaultmidpunct}
{\mcitedefaultendpunct}{\mcitedefaultseppunct}\relax
\EndOfBibitem
\bibitem[Palmieri \latin{et~al.}(2020)Palmieri, Kovlakov, Bianchi, Yudin,
  Straupe, Biamonte, and Kulik]{Palmieri2020}
Palmieri,~A.~M.; Kovlakov,~E.; Bianchi,~F.; Yudin,~D.; Straupe,~S.;
  Biamonte,~J.~D.; Kulik,~S. Experimental neural network enhanced quantum
  tomography. \emph{npj Quantum Information} \textbf{2020}, \emph{6}, 20\relax
\mciteBstWouldAddEndPuncttrue
\mciteSetBstMidEndSepPunct{\mcitedefaultmidpunct}
{\mcitedefaultendpunct}{\mcitedefaultseppunct}\relax
\EndOfBibitem
\bibitem[Xin \latin{et~al.}(2019)Xin, Lu, Cao, Anikeeva, Lu, Li, Long, and
  Zeng]{xin2019local}
Xin,~T.; Lu,~S.; Cao,~N.; Anikeeva,~G.; Lu,~D.; Li,~J.; Long,~G.; Zeng,~B.
  Local-measurement-based quantum state tomography via neural networks.
  \emph{npj Quantum Information} \textbf{2019}, \emph{5}, 1--8\relax
\mciteBstWouldAddEndPuncttrue
\mciteSetBstMidEndSepPunct{\mcitedefaultmidpunct}
{\mcitedefaultendpunct}{\mcitedefaultseppunct}\relax
\EndOfBibitem
\bibitem[Wichmann(1963)]{wich}
Wichmann,~E.~H. Density Matrices Arising from Incomplete Measurements. \emph{J.
  Math. Phys.} \textbf{1963}, \emph{4}, 884--896\relax
\mciteBstWouldAddEndPuncttrue
\mciteSetBstMidEndSepPunct{\mcitedefaultmidpunct}
{\mcitedefaultendpunct}{\mcitedefaultseppunct}\relax
\EndOfBibitem
\bibitem[Jaynes(1957)]{jaynes}
Jaynes,~E.~T. Information Theory and Statistical Mechanics. II. \emph{Phys.
  Rev.} \textbf{1957}, \emph{108}, 171--190\relax
\mciteBstWouldAddEndPuncttrue
\mciteSetBstMidEndSepPunct{\mcitedefaultmidpunct}
{\mcitedefaultendpunct}{\mcitedefaultseppunct}\relax
\EndOfBibitem
\bibitem[Katz(1967)]{katz1967principles}
Katz,~A. \emph{Principles of Statistical Mechanics: The Information Theory
  Approach}; W. H. Freeman, 1967\relax
\mciteBstWouldAddEndPuncttrue
\mciteSetBstMidEndSepPunct{\mcitedefaultmidpunct}
{\mcitedefaultendpunct}{\mcitedefaultseppunct}\relax
\EndOfBibitem
\bibitem[Sajjan \latin{et~al.}(2021)Sajjan, Li, Selvarajan, Sureshbabu, Kale,
  Gupta, and Kais]{sajjan2021quantum}
Sajjan,~M.; Li,~J.; Selvarajan,~R.; Sureshbabu,~S.~H.; Kale,~S.~S.; Gupta,~R.;
  Kais,~S. Quantum computing enhanced machine learning for physico-chemical
  applications. \emph{arXiv preprint arXiv:2111.00851} \textbf{2021}, \relax
\mciteBstWouldAddEndPunctfalse
\mciteSetBstMidEndSepPunct{\mcitedefaultmidpunct}
{}{\mcitedefaultseppunct}\relax
\EndOfBibitem
\bibitem[Gupta \latin{et~al.}(2021)Gupta, Xia, Levine, and Kais]{rishabh}
Gupta,~R.; Xia,~R.; Levine,~R.~D.; Kais,~S. Maximal Entropy Approach for
  Quantum State Tomography. \emph{PRX Quantum} \textbf{2021}, \emph{2},
  010318\relax
\mciteBstWouldAddEndPuncttrue
\mciteSetBstMidEndSepPunct{\mcitedefaultmidpunct}
{\mcitedefaultendpunct}{\mcitedefaultseppunct}\relax
\EndOfBibitem
\bibitem[Gupta \latin{et~al.}(2021)Gupta, Levine, and
  Kais]{gupta2021convergence}
Gupta,~R.; Levine,~R.~D.; Kais,~S. Convergence of a Reconstructed Density
  Matrix to a Pure State Using the Maximal Entropy Approach. \emph{The Journal
  of Physical Chemistry A} \textbf{2021}, \emph{125}, 7588--7595\relax
\mciteBstWouldAddEndPuncttrue
\mciteSetBstMidEndSepPunct{\mcitedefaultmidpunct}
{\mcitedefaultendpunct}{\mcitedefaultseppunct}\relax
\EndOfBibitem
\bibitem[Levine and Tribus(1979)Levine, and Tribus]{raphy}
Levine,~R.~D.; Tribus,~M. \emph{Maximum Entropy Formalism}; Cambridge, Mass. :
  MIT Press: USA, 1979\relax
\mciteBstWouldAddEndPuncttrue
\mciteSetBstMidEndSepPunct{\mcitedefaultmidpunct}
{\mcitedefaultendpunct}{\mcitedefaultseppunct}\relax
\EndOfBibitem
\bibitem[{Agmon} \latin{et~al.}(1979){Agmon}, {Alhassid}, and {Levine}]{raphy2}
{Agmon},~N.; {Alhassid},~Y.; {Levine},~R.~D. {An Algorithm for Finding the
  Distribution of Maximal Entropy}. \emph{J. Comput. Phys.} \textbf{1979},
  \emph{30}, 250--258\relax
\mciteBstWouldAddEndPuncttrue
\mciteSetBstMidEndSepPunct{\mcitedefaultmidpunct}
{\mcitedefaultendpunct}{\mcitedefaultseppunct}\relax
\EndOfBibitem
\bibitem[Bužek \latin{et~al.}(1997)Bužek, Drobný, Adam, Derka, and
  Knight]{Buzek_1997}
Bužek,~V.; Drobný,~G.; Adam,~G.; Derka,~R.; Knight,~P.~L. Reconstruction of
  quantum states of spin systems via the Jaynes principle of maximum entropy.
  \emph{Journal of Modern Optics} \textbf{1997}, \emph{44}, 2607–2627\relax
\mciteBstWouldAddEndPuncttrue
\mciteSetBstMidEndSepPunct{\mcitedefaultmidpunct}
{\mcitedefaultendpunct}{\mcitedefaultseppunct}\relax
\EndOfBibitem
\bibitem[Alhassid and Levine(1978)Alhassid, and Levine]{alhassid}
Alhassid,~Y.; Levine,~R.~D. Connection between the maximal entropy and the
  scattering theoretic analyses of collision processes. \emph{Phys. Rev. A}
  \textbf{1978}, \emph{18}, 89--116\relax
\mciteBstWouldAddEndPuncttrue
\mciteSetBstMidEndSepPunct{\mcitedefaultmidpunct}
{\mcitedefaultendpunct}{\mcitedefaultseppunct}\relax
\EndOfBibitem
\bibitem[Dagan and Dothan(1982)Dagan, and Dothan]{dagan}
Dagan,~S.; Dothan,~Y. Evaluation of an incompletely measured spin density
  matrix. \emph{Phys. Rev. D} \textbf{1982}, \emph{26}, 248--260\relax
\mciteBstWouldAddEndPuncttrue
\mciteSetBstMidEndSepPunct{\mcitedefaultmidpunct}
{\mcitedefaultendpunct}{\mcitedefaultseppunct}\relax
\EndOfBibitem
\bibitem[Lewin \latin{et~al.}(2021)Lewin, Nam, and
  Rougerie]{lewin2021classical}
Lewin,~M.; Nam,~P.~T.; Rougerie,~N. Classical field theory limit of many-body
  quantum Gibbs states in 2D and 3D. \emph{Inventiones mathematicae}
  \textbf{2021}, \emph{224}, 315--444\relax
\mciteBstWouldAddEndPuncttrue
\mciteSetBstMidEndSepPunct{\mcitedefaultmidpunct}
{\mcitedefaultendpunct}{\mcitedefaultseppunct}\relax
\EndOfBibitem
\bibitem[Childs \latin{et~al.}(2018)Childs, Maslov, Nam, Ross, and
  Su]{childs2018toward}
Childs,~A.~M.; Maslov,~D.; Nam,~Y.; Ross,~N.~J.; Su,~Y. Toward the first
  quantum simulation with quantum speedup. \emph{Proceedings of the National
  Academy of Sciences} \textbf{2018}, \emph{115}, 9456--9461\relax
\mciteBstWouldAddEndPuncttrue
\mciteSetBstMidEndSepPunct{\mcitedefaultmidpunct}
{\mcitedefaultendpunct}{\mcitedefaultseppunct}\relax
\EndOfBibitem
\bibitem[Somma \latin{et~al.}(2008)Somma, Boixo, Barnum, and
  Knill]{somma2008quantum}
Somma,~R.~D.; Boixo,~S.; Barnum,~H.; Knill,~E. Quantum simulations of classical
  annealing processes. \emph{Physical review letters} \textbf{2008},
  \emph{101}, 130504\relax
\mciteBstWouldAddEndPuncttrue
\mciteSetBstMidEndSepPunct{\mcitedefaultmidpunct}
{\mcitedefaultendpunct}{\mcitedefaultseppunct}\relax
\EndOfBibitem
\bibitem[Kieferov{\'a} and Wiebe(2017)Kieferov{\'a}, and
  Wiebe]{kieferova2017tomography}
Kieferov{\'a},~M.; Wiebe,~N. Tomography and generative training with quantum
  Boltzmann machines. \emph{Physical Review A} \textbf{2017}, \emph{96},
  062327\relax
\mciteBstWouldAddEndPuncttrue
\mciteSetBstMidEndSepPunct{\mcitedefaultmidpunct}
{\mcitedefaultendpunct}{\mcitedefaultseppunct}\relax
\EndOfBibitem
\bibitem[Biamonte \latin{et~al.}(2017)Biamonte, Wittek, Pancotti, Rebentrost,
  Wiebe, and Lloyd]{biamonte2017quantum}
Biamonte,~J.; Wittek,~P.; Pancotti,~N.; Rebentrost,~P.; Wiebe,~N.; Lloyd,~S.
  Quantum machine learning. \emph{Nature} \textbf{2017}, \emph{549},
  195--202\relax
\mciteBstWouldAddEndPuncttrue
\mciteSetBstMidEndSepPunct{\mcitedefaultmidpunct}
{\mcitedefaultendpunct}{\mcitedefaultseppunct}\relax
\EndOfBibitem
\bibitem[Wang \latin{et~al.}(2021)Wang, Li, and Wang]{wang2021variational}
Wang,~Y.; Li,~G.; Wang,~X. Variational quantum Gibbs state preparation with a
  truncated Taylor series. \emph{Physical Review Applied} \textbf{2021},
  \emph{16}, 054035\relax
\mciteBstWouldAddEndPuncttrue
\mciteSetBstMidEndSepPunct{\mcitedefaultmidpunct}
{\mcitedefaultendpunct}{\mcitedefaultseppunct}\relax
\EndOfBibitem
\bibitem[Poulin and Wocjan(2009)Poulin, and Wocjan]{poulin2009sampling}
Poulin,~D.; Wocjan,~P. Sampling from the thermal quantum Gibbs state and
  evaluating partition functions with a quantum computer. \emph{Physical review
  letters} \textbf{2009}, \emph{103}, 220502\relax
\mciteBstWouldAddEndPuncttrue
\mciteSetBstMidEndSepPunct{\mcitedefaultmidpunct}
{\mcitedefaultendpunct}{\mcitedefaultseppunct}\relax
\EndOfBibitem
\bibitem[Temme \latin{et~al.}(2011)Temme, Osborne, Vollbrecht, Poulin, and
  Verstraete]{temme2011quantum}
Temme,~K.; Osborne,~T.~J.; Vollbrecht,~K.~G.; Poulin,~D.; Verstraete,~F.
  Quantum metropolis sampling. \emph{Nature} \textbf{2011}, \emph{471},
  87--90\relax
\mciteBstWouldAddEndPuncttrue
\mciteSetBstMidEndSepPunct{\mcitedefaultmidpunct}
{\mcitedefaultendpunct}{\mcitedefaultseppunct}\relax
\EndOfBibitem
\bibitem[Kastoryano and Brandao(2016)Kastoryano, and
  Brandao]{kastoryano2016quantum}
Kastoryano,~M.~J.; Brandao,~F.~G. Quantum Gibbs samplers: The commuting case.
  \emph{Communications in Mathematical Physics} \textbf{2016}, \emph{344},
  915--957\relax
\mciteBstWouldAddEndPuncttrue
\mciteSetBstMidEndSepPunct{\mcitedefaultmidpunct}
{\mcitedefaultendpunct}{\mcitedefaultseppunct}\relax
\EndOfBibitem
\bibitem[Brand{\~a}o and Kastoryano(2019)Brand{\~a}o, and
  Kastoryano]{brandao2019finite}
Brand{\~a}o,~F.~G.; Kastoryano,~M.~J. Finite correlation length implies
  efficient preparation of quantum thermal states. \emph{Communications in
  Mathematical Physics} \textbf{2019}, \emph{365}, 1--16\relax
\mciteBstWouldAddEndPuncttrue
\mciteSetBstMidEndSepPunct{\mcitedefaultmidpunct}
{\mcitedefaultendpunct}{\mcitedefaultseppunct}\relax
\EndOfBibitem
\bibitem[Yung and Aspuru-Guzik(2012)Yung, and Aspuru-Guzik]{yung2012quantum}
Yung,~M.-H.; Aspuru-Guzik,~A. A quantum--quantum Metropolis algorithm.
  \emph{Proceedings of the National Academy of Sciences} \textbf{2012},
  \emph{109}, 754--759\relax
\mciteBstWouldAddEndPuncttrue
\mciteSetBstMidEndSepPunct{\mcitedefaultmidpunct}
{\mcitedefaultendpunct}{\mcitedefaultseppunct}\relax
\EndOfBibitem
\bibitem[Kaplan \latin{et~al.}(2017)Kaplan, Klco, and
  Roggero]{kaplan2017ground}
Kaplan,~D.~B.; Klco,~N.; Roggero,~A. Ground states via spectral combing on a
  quantum computer. \emph{arXiv preprint arXiv:1709.08250} \textbf{2017},
  \relax
\mciteBstWouldAddEndPunctfalse
\mciteSetBstMidEndSepPunct{\mcitedefaultmidpunct}
{}{\mcitedefaultseppunct}\relax
\EndOfBibitem
\bibitem[Riera \latin{et~al.}(2012)Riera, Gogolin, and
  Eisert]{riera2012thermalization}
Riera,~A.; Gogolin,~C.; Eisert,~J. Thermalization in nature and on a quantum
  computer. \emph{Physical review letters} \textbf{2012}, \emph{108},
  080402\relax
\mciteBstWouldAddEndPuncttrue
\mciteSetBstMidEndSepPunct{\mcitedefaultmidpunct}
{\mcitedefaultendpunct}{\mcitedefaultseppunct}\relax
\EndOfBibitem
\bibitem[Bilgin and Boixo(2010)Bilgin, and Boixo]{bilgin2010preparing}
Bilgin,~E.; Boixo,~S. Preparing thermal states of quantum systems by dimension
  reduction. \emph{Physical review letters} \textbf{2010}, \emph{105},
  170405\relax
\mciteBstWouldAddEndPuncttrue
\mciteSetBstMidEndSepPunct{\mcitedefaultmidpunct}
{\mcitedefaultendpunct}{\mcitedefaultseppunct}\relax
\EndOfBibitem
\bibitem[McClean \latin{et~al.}(2016)McClean, Romero, Babbush, and
  Aspuru-Guzik]{mcclean2016theory}
McClean,~J.~R.; Romero,~J.; Babbush,~R.; Aspuru-Guzik,~A. The theory of
  variational hybrid quantum-classical algorithms. \emph{New Journal of
  Physics} \textbf{2016}, \emph{18}, 023023\relax
\mciteBstWouldAddEndPuncttrue
\mciteSetBstMidEndSepPunct{\mcitedefaultmidpunct}
{\mcitedefaultendpunct}{\mcitedefaultseppunct}\relax
\EndOfBibitem
\bibitem[Bravo-Prieto \latin{et~al.}(2019)Bravo-Prieto, LaRose, Cerezo, Subasi,
  Cincio, and Coles]{bravo2019variational}
Bravo-Prieto,~C.; LaRose,~R.; Cerezo,~M.; Subasi,~Y.; Cincio,~L.; Coles,~P.~J.
  Variational quantum linear solver. \emph{arXiv preprint arXiv:1909.05820}
  \textbf{2019}, \relax
\mciteBstWouldAddEndPunctfalse
\mciteSetBstMidEndSepPunct{\mcitedefaultmidpunct}
{}{\mcitedefaultseppunct}\relax
\EndOfBibitem
\bibitem[Huang \latin{et~al.}(2019)Huang, Bharti, and
  Rebentrost]{huang2019near}
Huang,~H.-Y.; Bharti,~K.; Rebentrost,~P. Near-term quantum algorithms for
  linear systems of equations. \emph{arXiv preprint arXiv:1909.07344}
  \textbf{2019}, \relax
\mciteBstWouldAddEndPunctfalse
\mciteSetBstMidEndSepPunct{\mcitedefaultmidpunct}
{}{\mcitedefaultseppunct}\relax
\EndOfBibitem
\bibitem[LaRose \latin{et~al.}(2019)LaRose, Tikku, O’Neel-Judy, Cincio, and
  Coles]{larose2019variational}
LaRose,~R.; Tikku,~A.; O’Neel-Judy,~{\'E}.; Cincio,~L.; Coles,~P.~J.
  Variational quantum state diagonalization. \emph{npj Quantum Information}
  \textbf{2019}, \emph{5}, 1--10\relax
\mciteBstWouldAddEndPuncttrue
\mciteSetBstMidEndSepPunct{\mcitedefaultmidpunct}
{\mcitedefaultendpunct}{\mcitedefaultseppunct}\relax
\EndOfBibitem
\bibitem[Cerezo \latin{et~al.}(2020)Cerezo, Sharma, Arrasmith, and
  Coles]{cerezo2020variational}
Cerezo,~M.; Sharma,~K.; Arrasmith,~A.; Coles,~P.~J. Variational quantum state
  eigensolver. \emph{arXiv preprint arXiv:2004.01372} \textbf{2020}, \relax
\mciteBstWouldAddEndPunctfalse
\mciteSetBstMidEndSepPunct{\mcitedefaultmidpunct}
{}{\mcitedefaultseppunct}\relax
\EndOfBibitem
\bibitem[Peruzzo \latin{et~al.}(2014)Peruzzo, McClean, Shadbolt, Yung, Zhou,
  Love, Aspuru-Guzik, and O’brien]{peruzzo2014variational}
Peruzzo,~A.; McClean,~J.; Shadbolt,~P.; Yung,~M.-H.; Zhou,~X.-Q.; Love,~P.~J.;
  Aspuru-Guzik,~A.; O’brien,~J.~L. A variational eigenvalue solver on a
  photonic quantum processor. \emph{Nature communications} \textbf{2014},
  \emph{5}, 1--7\relax
\mciteBstWouldAddEndPuncttrue
\mciteSetBstMidEndSepPunct{\mcitedefaultmidpunct}
{\mcitedefaultendpunct}{\mcitedefaultseppunct}\relax
\EndOfBibitem
\bibitem[Chen \latin{et~al.}(2021)Chen, Song, Zhao, and
  Wang]{chen2021variational}
Chen,~R.; Song,~Z.; Zhao,~X.; Wang,~X. Variational quantum algorithms for trace
  distance and fidelity estimation. \emph{Quantum Science and Technology}
  \textbf{2021}, \emph{7}, 015019\relax
\mciteBstWouldAddEndPuncttrue
\mciteSetBstMidEndSepPunct{\mcitedefaultmidpunct}
{\mcitedefaultendpunct}{\mcitedefaultseppunct}\relax
\EndOfBibitem
\bibitem[Islam \latin{et~al.}(2015)Islam, Ma, Preiss, Eric~Tai, Lukin, Rispoli,
  and Greiner]{islam2015measuring}
Islam,~R.; Ma,~R.; Preiss,~P.~M.; Eric~Tai,~M.; Lukin,~A.; Rispoli,~M.;
  Greiner,~M. Measuring entanglement entropy in a quantum many-body system.
  \emph{Nature} \textbf{2015}, \emph{528}, 77--83\relax
\mciteBstWouldAddEndPuncttrue
\mciteSetBstMidEndSepPunct{\mcitedefaultmidpunct}
{\mcitedefaultendpunct}{\mcitedefaultseppunct}\relax
\EndOfBibitem
\bibitem[Verdon \latin{et~al.}(2019)Verdon, Marks, Nanda, Leichenauer, and
  Hidary]{verdon2019quantum}
Verdon,~G.; Marks,~J.; Nanda,~S.; Leichenauer,~S.; Hidary,~J. Quantum
  Hamiltonian-based models and the variational quantum thermalizer algorithm.
  \emph{arXiv preprint arXiv:1910.02071} \textbf{2019}, \relax
\mciteBstWouldAddEndPunctfalse
\mciteSetBstMidEndSepPunct{\mcitedefaultmidpunct}
{}{\mcitedefaultseppunct}\relax
\EndOfBibitem
\bibitem[Wu and Hsieh(2019)Wu, and Hsieh]{wu2019variational}
Wu,~J.; Hsieh,~T.~H. Variational thermal quantum simulation via thermofield
  double states. \emph{Physical review letters} \textbf{2019}, \emph{123},
  220502\relax
\mciteBstWouldAddEndPuncttrue
\mciteSetBstMidEndSepPunct{\mcitedefaultmidpunct}
{\mcitedefaultendpunct}{\mcitedefaultseppunct}\relax
\EndOfBibitem
\bibitem[McArdle \latin{et~al.}(2019)McArdle, Jones, Endo, Li, Benjamin, and
  Yuan]{mcardle2019variational}
McArdle,~S.; Jones,~T.; Endo,~S.; Li,~Y.; Benjamin,~S.~C.; Yuan,~X. Variational
  ansatz-based quantum simulation of imaginary time evolution. \emph{npj
  Quantum Information} \textbf{2019}, \emph{5}, 1--6\relax
\mciteBstWouldAddEndPuncttrue
\mciteSetBstMidEndSepPunct{\mcitedefaultmidpunct}
{\mcitedefaultendpunct}{\mcitedefaultseppunct}\relax
\EndOfBibitem
\bibitem[Yuan \latin{et~al.}(2019)Yuan, Endo, Zhao, Li, and
  Benjamin]{yuan2019theory}
Yuan,~X.; Endo,~S.; Zhao,~Q.; Li,~Y.; Benjamin,~S.~C. Theory of variational
  quantum simulation. \emph{Quantum} \textbf{2019}, \emph{3}, 191\relax
\mciteBstWouldAddEndPuncttrue
\mciteSetBstMidEndSepPunct{\mcitedefaultmidpunct}
{\mcitedefaultendpunct}{\mcitedefaultseppunct}\relax
\EndOfBibitem
\bibitem[Chowdhury \latin{et~al.}(2020)Chowdhury, Low, and
  Wiebe]{chowdhury2020variational}
Chowdhury,~A.~N.; Low,~G.~H.; Wiebe,~N. A variational quantum algorithm for
  preparing quantum Gibbs states. \emph{arXiv preprint arXiv:2002.00055}
  \textbf{2020}, \relax
\mciteBstWouldAddEndPunctfalse
\mciteSetBstMidEndSepPunct{\mcitedefaultmidpunct}
{}{\mcitedefaultseppunct}\relax
\EndOfBibitem
\bibitem[Abraham \latin{et~al.}(2019)Abraham, AduOffei, Agarwal, Akhalwaya,
  Aleksandrowicz, Alexander, Arbel, Asfaw, Azaustre, AzizNgoueya, Bansal,
  Barkoutsos, Barron, Bello, Ben-Haim, Bevenius, Bishop, Bolos, Bosch, Bravyi,
  Bucher, Burov, Cabrera, Calpin, Capelluto, Carballo, Carrascal, Chen, Chen,
  Chen, Chen, Chen, Chow, Churchill, Claus, Clauss, Cocking, Cross, Cross,
  Cross, Cruz-Benito, Culver, C{\'o}rcoles-Gonzales, Dague, Dandachi, Daniels,
  Dartiailh, DavideFrr, Davila, Dekusar, Ding, Doi, Drechsler, Drew,
  Dumitrescu, Dumon, Duran, EL-Safty, Eastman, Eendebak, Egger, Everitt,
  Fern{\'a}ndez, Ferrera, Fouilland, FranckChevallier, Frisch, Fuhrer, GEORGE,
  Gacon, Gago, Gambella, Gambetta, Gammanpila, Garcia, Garion, Gilliam,
  Giridharan, Gomez-Mosquera, de~la Puente~Gonz{\'a}lez, Gorzinski, Gould,
  Greenberg, Grinko, Guan, Gunnels, Haglund, Haide, Hamamura, Hamido, Havlicek,
  Hellmers, Herok, Hillmich, Horii, Howington, Hu, Hu, Huisman, Imai, Imamichi,
  Ishizaki, Iten, Itoko, JamesSeaward, Javadi, Javadi-Abhari, Jessica,
  Jivrajani, Johns, Jonathan-Shoemaker, Kachmann, Kanazawa, Kang-Bae, Karazeev,
  Kassebaum, King, Knabberjoe, Kobayashi, Kovyrshin, Krishnakumar, Krishnan,
  Krsulich, Kus, LaRose, Lacal, Lambert, Lapeyre, Latone, Lawrence, Lee, Li,
  Liu, Liu, Maeng, Malyshev, Manela, Marecek, Marques, Maslov, Mathews, Matsuo,
  McClure, McGarry, McKay, McPherson, Meesala, Metcalfe, Mevissen, Mezzacapo,
  Midha, Minev, Mitchell, Moll, Mooring, Morales, Moran, MrF, Murali,
  M{\"u}ggenburg, Nadlinger, Nakanishi, Nannicini, Nation, Navarro, Naveh,
  Neagle, Neuweiler, Niroula, Norlen, O'Riordan, Ogunbayo, Ollitrault, Oud,
  Padilha, Paik, Pang, Perriello, Phan, Piro, Pistoia, Piveteau,
  Pozas-iKerstjens, Prutyanov, Puzzuoli, P{\'e}rez, Quintiii, Rahman, Raja,
  Ramagiri, Rao, Raymond, Redondo, Reuter, Rice, Rodr{\'\i}guez, RohithKarur,
  Rossmannek, Ryu, SAPV, SamFerracin, Sandberg, Sapra, Sargsyan, Sarkar,
  Sathaye, Schmitt, Schnabel, Schoenfeld, Scholten, Schoute, Schwarm, Sertage,
  Setia, Shammah, Shi, Silva, Simonetto, Singstock, Siraichi, Sitdikov,
  Sivarajah, Sletfjerding, Smolin, Soeken, Sokolov, SooluThomas, Starfish,
  Steenken, Stypulkoski, Sun, Sung, Takahashi, Tavernelli, Taylor, Taylour,
  Thomas, Tillet, Tod, Tomasik, de~la Torre, Trabing, Treinish, TrishaPe,
  Turner, Vaknin, Valcarce, Varchon, Vazquez, Villar, Vogt-Lee, Vuillot,
  Weaver, Wieczorek, Wildstrom, Winston, Woehr, Woerner, Woo, Wood, Wood, Wood,
  Wood, Wootton, Yeralin, Yonge-Mallo, Young, Yu, Zachow, Zdanski, Zhang,
  Zoufal, Zoufalc, a~kapila, a~matsuo, bcamorrison, brandhsn, chlorophyll zz,
  dekel.meirom, dekool, dime10, drholmie, dtrenev, ehchen, elfrocampeador,
  faisaldebouni, fanizzamarco, gadial, gruu, hhorii, hykavitha, jagunther,
  jliu45, kanejess, klinvill, kurarrr, lerongil, ma5x, merav aharoni,
  michelle4654, ordmoj, rmoyard, saswati qiskit, sethmerkel, strickroman,
  sumitpuri, tigerjack, toural, vvilpas, welien, willhbang, yang.luh,
  yotamvakninibm, and {\v{C}}epulkovskis]{Qiskit}
Abraham,~H. \latin{et~al.}  Qiskit: An Open-source Framework for Quantum
  Computing. 2019\relax
\mciteBstWouldAddEndPuncttrue
\mciteSetBstMidEndSepPunct{\mcitedefaultmidpunct}
{\mcitedefaultendpunct}{\mcitedefaultseppunct}\relax
\EndOfBibitem
\bibitem[Kandala \latin{et~al.}(2017)Kandala, Mezzacapo, Temme, Takita, Brink,
  Chow, and Gambetta]{kandala2017hardware}
Kandala,~A.; Mezzacapo,~A.; Temme,~K.; Takita,~M.; Brink,~M.; Chow,~J.~M.;
  Gambetta,~J.~M. Hardware-efficient variational quantum eigensolver for small
  molecules and quantum magnets. \emph{Nature} \textbf{2017}, \emph{549},
  242--246\relax
\mciteBstWouldAddEndPuncttrue
\mciteSetBstMidEndSepPunct{\mcitedefaultmidpunct}
{\mcitedefaultendpunct}{\mcitedefaultseppunct}\relax
\EndOfBibitem
\bibitem[Bian \latin{et~al.}(2019)Bian, Murphy, Xia, Daskin, and
  Kais]{bian2019quantum}
Bian,~T.; Murphy,~D.; Xia,~R.; Daskin,~A.; Kais,~S. Quantum computing methods
  for electronic states of the water molecule. \emph{Molecular Physics}
  \textbf{2019}, \emph{117}, 2069--2082\relax
\mciteBstWouldAddEndPuncttrue
\mciteSetBstMidEndSepPunct{\mcitedefaultmidpunct}
{\mcitedefaultendpunct}{\mcitedefaultseppunct}\relax
\EndOfBibitem
\bibitem[Gheorghiu and Hoban(2020)Gheorghiu, and
  Hoban]{gheorghiu2020estimating}
Gheorghiu,~A.; Hoban,~M.~J. Estimating the entropy of shallow circuit outputs
  is hard. \emph{arXiv preprint arXiv:2002.12814} \textbf{2020}, \relax
\mciteBstWouldAddEndPunctfalse
\mciteSetBstMidEndSepPunct{\mcitedefaultmidpunct}
{}{\mcitedefaultseppunct}\relax
\EndOfBibitem
\bibitem[Buhrman \latin{et~al.}(2001)Buhrman, Cleve, Watrous, and
  De~Wolf]{buhrman2001quantum}
Buhrman,~H.; Cleve,~R.; Watrous,~J.; De~Wolf,~R. Quantum fingerprinting.
  \emph{Physical Review Letters} \textbf{2001}, \emph{87}, 167902\relax
\mciteBstWouldAddEndPuncttrue
\mciteSetBstMidEndSepPunct{\mcitedefaultmidpunct}
{\mcitedefaultendpunct}{\mcitedefaultseppunct}\relax
\EndOfBibitem
\bibitem[Gottesman and Chuang(2001)Gottesman, and Chuang]{gottesman2001quantum}
Gottesman,~D.; Chuang,~I. Quantum digital signatures. \emph{arXiv preprint
  quant-ph/0105032} \textbf{2001}, \relax
\mciteBstWouldAddEndPunctfalse
\mciteSetBstMidEndSepPunct{\mcitedefaultmidpunct}
{}{\mcitedefaultseppunct}\relax
\EndOfBibitem
\bibitem[Patel \latin{et~al.}(2016)Patel, Ho, Ferreyrol, Ralph, and
  Pryde]{patel2016quantum}
Patel,~R.~B.; Ho,~J.; Ferreyrol,~F.; Ralph,~T.~C.; Pryde,~G.~J. A quantum
  Fredkin gate. \emph{Science advances} \textbf{2016}, \emph{2}, e1501531\relax
\mciteBstWouldAddEndPuncttrue
\mciteSetBstMidEndSepPunct{\mcitedefaultmidpunct}
{\mcitedefaultendpunct}{\mcitedefaultseppunct}\relax
\EndOfBibitem
\bibitem[Linke \latin{et~al.}(2018)Linke, Johri, Figgatt, Landsman, Matsuura,
  and Monroe]{linke2018measuring}
Linke,~N.~M.; Johri,~S.; Figgatt,~C.; Landsman,~K.~A.; Matsuura,~A.~Y.;
  Monroe,~C. Measuring the R{\'e}nyi entropy of a two-site Fermi-Hubbard model
  on a trapped ion quantum computer. \emph{Physical Review A} \textbf{2018},
  \emph{98}, 052334\relax
\mciteBstWouldAddEndPuncttrue
\mciteSetBstMidEndSepPunct{\mcitedefaultmidpunct}
{\mcitedefaultendpunct}{\mcitedefaultseppunct}\relax
\EndOfBibitem
\end{mcitethebibliography}

\end{document}